\documentclass[10pt, a4paper]{article}

\usepackage[final]{lrec2026} %
\usepackage{booktabs}
\usepackage{multicol}
\usepackage{multirow}
\usepackage{amsmath}
\usepackage{float}
\usepackage{soul}

\newcommand{\emoji}[1]{%
    \raisebox{-.15\baselineskip}{%
        \includegraphics[height=2ex,alt={#1 emoji}]{emojis/#1_google.png}%
    }%
}

\usepackage{tikz}
\usetikzlibrary{positioning, fit, shapes.multipart}

\title{Data Selection Effects on Self-Supervised Learning of Audio Representations for French Audiovisual Broadcasts}

\name{Valentin Pelloin$^*$, Lina Bekkali$^{*\dagger}$, Reda Dehak$^\ddagger$, David Doukhan$^*$} 

\address{
    $^*$Institut National de l'Audiovisuel (INA), France, \\
    $^\dagger$École nationale des ponts et chaussées (ENPC), France, \\
    $^\ddagger$EPITA Research Laboratory (LRE), France \\
         \{vpelloin, ddoukhan\}@ina.fr, {lina.bekkali@eleves.enpc.fr}, reda.dehak@epita.fr\\
         }

\abstract{
Audio and speech self-supervised encoder models are now widely used for a lot of different tasks. Many of these models are often trained on clean segmented speech content such as LibriSpeech. In this paper, we look into how the pretraining datasets of such SSL (Self-Supervised Learning) models impact their downstream results. We build a large pretraining corpus of highly diverse TV and Radio broadcast audio content, which we describe with automatic tools. We use these annotations to build smaller subsets, which we use to train audio SSL models. Then, we evaluate the models on multiple downstream tasks such as automatic speech recognition, voice activity and music detection, or speaker recognition.
The results show the potential of pretraining SSL models on diverse audio content without restricting it to speech.
We also perform a membership inference attack to evaluate the encoder ability to memorize their training datasets, which highlight the importance of data deduplication.
This unified training could bridge speech and music machine learning communities.
 \\ \newline }

\tikzset{
  transformerLayer/.style={
    draw, fill=gray!15, minimum width=3cm, minimum height=0.2cm, font={\footnotesize}, execute at begin node=\setlength{\baselineskip}{2pt}, inner sep=1.5pt
  },
  transformerLayerIgnored/.style={
    transformerLayer, draw=gray, fill=gray!5, text=gray
  },
  dataBlock/.style={
    draw, rounded corners, thick, inner sep=4pt, color=cyan!80, minimum width=2.5cm
  },
  processingBlock/.style={
    draw, rounded corners, thick, inner sep=4pt, color=gray, minimum width=2.5cm
  },
  autoBlock/.style={
    draw, rounded corners, thick, inner sep=2pt, outer sep=0pt, dotted,
  },
  arrowType/.style={
    ->, thick, color=gray
  },
}

\newcommand{\within}[1]{\setul{0.4mm}{}\setulcolor{black!50}\ul{#1}}

\newcommand{\base}{{\small{}\texttt{base}}}
\newcommand{\nomusic}{{\small{}\texttt{no\_music}}}
\newcommand{\onlyspeech}{{\small{}\texttt{only\_speech}}}
\newcommand{\onlyfr}{{\small{}\texttt{only\_fr}}}
\newcommand{\gender}{{\small{}\texttt{gender}}}
\newcommand{\duplicates}{{\small{}\texttt{duplicates}}}

\begin{document}

\maketitleabstract

\section{Introduction}
    Self-Supervised Learning (SSL) consists in pretraining models on unsupervised data, without using labeled data.
    In the context of audio and speech SSL models, an encoder model is pretrained on a large corpus of audio content. This model then generates embeddings that can be finetuned and used as input of downstream models to perform various tasks: music information retrieval, automatic speech recognition, speaker recognition, etc.
    However, speech encoders are trained on clean segmented speech \citep{lebenchmark2,zanonboito24_interspeech}, with many of them using read or audiobooks datasets as LibriSpeech \citep{wav2vec2,baevski2023efficient}.
    Having access to large corpora of audio content without any clean annotation or segmentation of speech, one might be tempted to pretrain an audio encoder on this content. However, there remain multiple questions on the viability of such approaches.
    If this model is pretrained on content which includes music, noises, and speech, will it obtain good performances on downstream tasks such as speech recognition?
    Will the generated features be useful to perform voice activity or music detection?

    Gender biases are commonly present in speech models \cite{addadecker05_interspeech,attanasio-etal-2024-twists,genderasr}, including in SSL models \citep{biasasrw2v,zanonboito22_interspeech}. Does balancing the pretraining data across speaker genders reduce this bias for downstream tasks such as ASR or speaker recognition?

    Another question arises from the quantity of duplicated content inside the pretraining dataset. For Natural Language Processing, others \cite{lee-etal-2022-deduplicating,CarliniExtracting} have demonstrated the negative impact it has on performances and sensitive data extraction. However, this impact has yet to be demonstrated on speech and audio tasks.
    As we release the pretrained and downstream models, we want to hinder the ability to extract pretraining data information from the model.

    In this paper, we introduce a new 100,000 hours audio corpus derived from TV and Radio broadcasts from \textit{Institut National de l'Audiovisuel} (INA, the French National Audiovisual Institute).
    This corpus is deduplicated using an audio deduplication tool, and segments are automatically described.
    With these pieces of information, we construct and pretrain 6 different audio SSL models, each on a subsample of 1,000 hours of content.
    To answer the questions highlighted above, we construct each subsample in order to evaluate the consequences of data selection during pretraining on the downstream evaluations.
    We evaluate our models on multiple downstream tasks: (gendered) automatic speech recognition, voice activity detection, music detection, speaker recognition, and a membership inference attack.

    Pretrained and downstream models are published on HuggingFace\footnote{\url{https://hf.co/spaces/ina-foss/LREC-2026-Data-Selection-Effects}}.
    Due to obvious copyright concerns, the training datasets are not released to the public,
    however, researchers seeking to obtain audiovisual archives may address their requests to \textit{Le Lab}, an entity dedicated to
    researchers willing to access French audiovisual archives~\citeplanguageresource{inatheque}.

\section{Related works} %
    Introduced by \citet{baevski2023efficient}, data2vec2 is an efficient and multimodal architecture to train SSL encoders. This architecture is composed of a teacher-student encoder, and can be trained with similar objectives for text, image or speech.
    Using an equivalent architecture, \citet{li2022mapmusic2vecsimpleeffectivebaseline} published music2vec, a model pretrained on music (1,000h). It obtains SOTA results on multiple Music Information Extraction (MIR) tasks.
    However, there does not exist a single unified encoder model suitable for both French speech and non speech tasks.

    \citet{lebenchmark2} released pretrained speech encoders models for French following the architecture of Wav2vec 2.0 \citep{wav2vec2}. 
    This architecture was also employed by \citet{zanonboito22_interspeech}, where they evaluated the impact of pretraining gender biases for the downstream ASR task, suggesting that gender-balanced pretraining might provide a better initialization for the finetuning process.

    Voice Activity Detection (VAD) using self-supervised speech representations has been experimented with success by \citet{gimeno21_interspeech,10094972,karan24_interspeech}. In particular, \citet{karan24_interspeech} used a Wav2vec~2.0 encoder pretrained on 436k hours of speech and finetuned it alongside their downstream model to obtain SOTA performances, while keeping the throughput speed reasonable.

    Although VAD and Music detection are closely related, unified architectures and techniques have not been developed between the two communities. While \citet{doukhan_inaspeechsegmenter} presented a model that detects voice and music, it cannot do both at the same time, for example when someone speaks over background music.
    This limitation might arise from available corpora labelled with both information. To the best of our knowledge, only AVA-Speech \citeplanguageresource{chaudhuri18_interspeech} is labelled for both speech and music detection.

    The most recent Speaker Recognition (SR) models are based on large SSL models \citep{sanyuan22_stsp, novoselov23_interspeech,zhengyang22_icassp,peng23_slt,peng24_interspeech}. These models use a backend model to aggregate and map hidden and temporal representations onto an embedding representation for speaker recognition purposes. The cosine distance between the embeddings is then used as a scoring method. Various approaches have been proposed. In simple methods, representations from all hidden layers of the pretrained SSL model are averaged with learnable weights and then fed as input features to a standard speaker recognition model such as ECAPA-TDNN \citep{zhengyang22_icassp}. \citet{ novoselov23_interspeech} suggests using a TDNN-based backend to directly aggregate hidden Wav2vec 2.0 representations. \citet{peng23_slt} proposed an attention-based backend that uses key and value flow; the embeddings are then obtained via a weighted average. While these models demonstrate the importance of initial layers in defining the speaker embedding space \citep{ sanyuan22_stsp}, the best performance relies on finetuning the pretrained model.

    The ability of SSL models to remember (memorize) their training data has been studied by many. For example, for text models, \citet{CarliniExtracting} managed to generate URLs seen during the training of GPT-2 XL.
    For speech SSL models, \citet{tseng22_interspeech} successfully set up Membership Inference Attacks (MIA) where they probe the model representations for the memorization of speaker and utterance information.

\section{Audio datasets for SSL}
    \label{sec:audio_datasets}

    Our objective is to build audio SSL models as general as possible. 
    These models could work for both speech analysis tasks such as speech recognition, speech understanding, speaker diarization or verification; and also for Music Information Retrieval (MIR) tasks: music and singing voice detection.
    We aim at applying these models on audiovisual archives to extract audio embeddings, and use these as input of multiple downstream classifiers which could describe content at scale automatically.

    \begin{figure}[htb!]
        \centering
        \includegraphics[width=\linewidth]{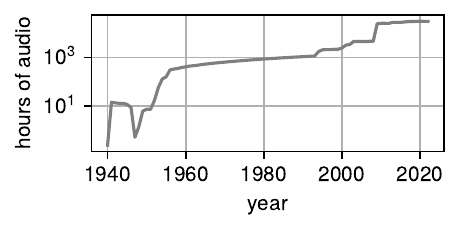}
        \vspace{-1cm}
        \caption{Number of hours of audio content in the source INA dataset per year. }
        \label{fig:duration_per_year}
    \end{figure}
    
    The \textit{Institut National de l'Audiovisuel} (INA) is in charge since 1975 of collecting and archiving TV and Radio content in France.
    In partnership with them, we obtained %
    a randomly sampled dataset of 473k hours of content, broadcast on 113 French TV and Radio channels, from 1940 to 2022. Thus, this dataset covers various kinds of audiovisual content: news, adverts, documentaries, game shows, movies, musics, cartoons, sports, etc. The dataset is composed of mostly 1h long audio files, each file corresponding to an unsegmented chunk of broadcast content on one channel on a particular date.
    In Figure \ref{fig:duration_per_year} the number of audio content in this dataset each year is presented.
    Since 1995, new legislations are enforcing archival of TV and Radio (legal deposit). 
    We notice a large increment starting in the recent years, thanks to the legal deposit\footnote{Legal deposit content between 1995 and 2009 was not broadly available due to storage format constraints.}. %

    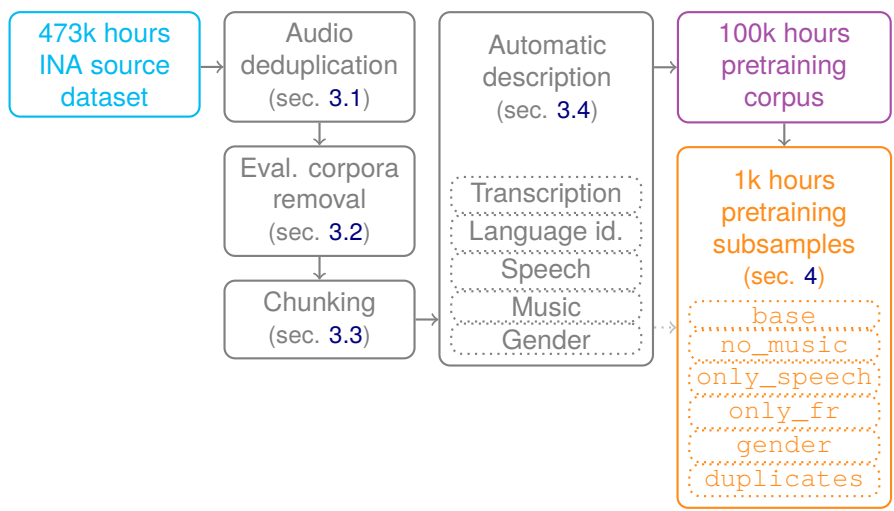
\begin{figure*}[htbp!]
    \centering
\begin{tikzpicture}[font=\sffamily, every node/.style={align=center},node distance = 0.3cm]

\node[dataBlock] (ina_source) {473k hours\\INA source\\dataset};
\node[processingBlock, right=of ina_source.north east, anchor=north west] (dedup) {Audio\\deduplication\\{\small(sec. \ref{sec:audio_dedup})}};
\node[processingBlock, below=of dedup] (eval) {Eval. corpora\\removal\\{\small(sec. \ref{sec:removal_eval})}};
\node[processingBlock, below=of eval] (chunking) {Chunking\\{\small(sec. \ref{sec:chunking})}};
\node [matrix,processingBlock, right=of dedup.north east, anchor=north west] (auto)
  {
    \node[] {Automatic\\description\\{\small(sec. \ref{sec:auto_description})}}; \\
    \node[minimum height=0.55cm] {}; \\
    \node[autoBlock] {Transcription}; \\
    \node[autoBlock] {Language id.}; \\
    \node[autoBlock] {Speech}; \\
    \node[autoBlock] {Music}; \\
    \node[autoBlock] {Gender}; \\
  };
\node[dataBlock, color=violet!70, right=of auto.north east, anchor=north west, minimum width=2.8cm] (100k) {100k hours\\pretraining\\corpus};

\node [matrix,dataBlock, color=orange!90, below=of 100k] (subsets)
  {
    \node[] {1k hours\\pretraining\\subsamples\\{\small(sec. \ref{sec:ssl})}}; \\
    \node[autoBlock] {\texttt{base}}; \\
    \node[autoBlock] {\texttt{no\_music}}; \\
    \node[autoBlock] {\texttt{only\_speech}}; \\
    \node[autoBlock] {\texttt{only\_fr}}; \\
    \node[autoBlock] {\texttt{gender}}; \\
    \node[autoBlock] {\texttt{duplicates}}; \\
  };

\draw[arrowType] (ina_source.east |- dedup.west) -- (dedup);
\draw[arrowType] (dedup) -- (eval);
\draw[arrowType] (eval) -- (chunking);
\draw[arrowType] (chunking) -- (auto.west |- chunking.east);
\draw[arrowType] (auto.east |- 100k.west) -- (100k);
\draw[arrowType] (100k) -- (subsets);
\draw[arrowType, dotted, draw=gray!50] (auto.east |- subsets.west) -- (subsets);

\end{tikzpicture}
\caption{Overview of the data preprocessing pipeline.}
    \label{fig:preprocessing}
\end{figure*}

    We present in Figure \ref{fig:preprocessing} an overview of the data preprocessing pipeline, that we describe in the following sections.

    \subsection{Audio deduplication}
        \label{sec:audio_dedup}
        \citet{chenot:hal-01017118} showed that there was on average only about 8h of fresh content every day across 12 French TV channels, with some of them having as low as 2.3h of fresh content on each day.
        Many noted that deduplication is an important preprocessing step when preparing datasets for machine learning: it allows for faster training and better generalisation \cite{mikolov-etal-2018-advances,lee-etal-2022-deduplicating}.
        A more serious issue regarding data privacy was raised by others \citep{CarliniExtracting,pmlr-v162-kandpal22a,yan2024protectingdataprivacylarge}. \citet{CarliniExtracting} shows that the more sensitive information is repeated, the more it is at risk for memorization. According to \citet{pmlr-v162-kandpal22a}, text sequences present 10 times in training data of Language Models are on average generated 1000x more often that sequences present only once.

        In order to mitigate this risk, we decided to deduplicate our training dataset using the repeated content detection tool described by \citet{chenot:hal-01017118}.
        The tool extracts lightweight audio fingerprints.
        A database of all fingerprints is constructed, 
        copies are detected when at least 4 similar consecutive fingerprints are found,
        and we discard all copies of a content once it has already been found.
        The tool is described to be robust to many signal alterations, such as low pass filtering or temporal splits into short extracts (98\% recall with 24s chunks).
        154k hours of audio content were removed from our corpus with this deduplication step, representing 32.6\% of the original corpus.

    \subsection{Removal of evaluation corpora}
        \label{sec:removal_eval}
        We intend to evaluate our models on corpora containing French audiovisual contents. Therefore, we want to avoid pretraining on content found in evaluation datasets.
        As in section \ref{sec:audio_dedup}, we use \citeauthor{chenot:hal-01017118}'s tool, this time to remove existing datasets from our training corpus.
        We first aggregate the fingerprints of multiple datasets, including:
        ESTER1~\citelanguageresource{gravier-etal-2004-ester},
        ESTER2~\citelanguageresource{galliano09_interspeech},
        EPAC~\citelanguageresource{esteve-etal-2010-epac},
        QUAERO~\citelanguageresource{boudahmane-etal-2011-advances},
        ETAPE~\citelanguageresource{gravier-etal-2012-etape},
        REPERE~\citelanguageresource{giraudel-etal-2012-repere},
        Rhapsodie~\citelanguageresource{lacheret-etal-2014-rhapsodie},
        Orféo~\citelanguageresource{benzitoun2016projet},
        InaGVAD~\citelanguageresource{doukhan-etal-2024-inagvad},
        is24\_news\_topic~\citelanguageresource{pelloin24_interspeech}.
        We then remove all audio chunks that matches with these fingerprints.
        623h were removed from the 473k hours dataset in this step.

    \subsection{Chunking}
        \label{sec:chunking}
        Finally, we randomly sample 12M audio chunks of 30s, in order to create a corpus of 100,000 hours from the remaining content available. This step is necessary to restrict the required processing time for the automatic content description tools presented in the next section, while also ensuring data diversity.
        As a result of audio deduplication and chunking, this corpus of 100k hours represents 21\% of the original corpus provided by INA.

    \subsection{Automatic description of audio chunks}
        \label{sec:auto_description}
        We use different tools to automatically describe the 12M audio chunks. These data
        are then used in sec. \ref{sec:ssl}
        to obtain controlled pretraining corpora.
        
        We first use Whisper (\textit{\small{}whisper-large-v3-turbo}, \citealp{pmlr-v202-radford23a}) to transcribe audio chunks into text. We do not set the content language and let Whisper perform the language identification.

        We use InaSpeechSegmenter \cite{doukhan_inaspeechsegmenter}, a Voice Activity Detection (VAD) and Speaker Gender Segmentation (SGS) tool built for TV and radio audiovisual content. It has already been used by others as a dataset curation tool for SSL \cite{zanonboito24_interspeech}, to obtain clean speech segments. It allows us to predict a segmentation with active speech along with gender information. InaSpeechSegmenter also predicts a label ``music'' and ``noise'' but unfortunately, it cannot predict speech and music separately: it cannot tell if there is music in the background behind speech.

        Although many open-source VAD systems exist, we did not find music detection tools suitable for our needs.
        Instead,
        we bootstrap a small MLP model using embeddings generated by {music2vec} \cite{li2022mapmusic2vecsimpleeffectivebaseline} and finetuned on OpenBMAT \citelanguageresource{MelndezCataln2019} to predict the proportion of the \textit{no-music}, \textit{background-music} and \textit{foreground music} classes. 
        It obtains a Mean Absolute Error of 12.54\% globally across OpenBMAT (13.30\% on our test split).

    \subsection{Dataset statistics}
        \begin{table}[htb!]
            \centering
            \begin{tabular}{rl}
            \toprule
                \multicolumn{2}{c}{\textbf{Global:}} \\
                Segment duration & 30s \\
                Audio segments & 12,000,000 \\
                Total duration & 100,000h \\ 
                \multirow{2}{*}{Period} & [01/01/1940 \\
                & - 31/12/2022] \\
                Channels (TV+Radio) & 113 \\
                \midrule
                \multicolumn{2}{c}{\textbf{Content type$^*$:}} \\
                Segments with speech & 72.51\% \\ 
                Segments with music & 55.23\% \\  \midrule
                \multicolumn{2}{c}{\textbf{Gender balance$^*$:}} \\
                Women speaking time & 29.95\% \\
                Men speaking time & 70.05\% \\ \midrule
                \multicolumn{2}{c}{\textbf{Language$^*$:}} \\
                \underline{French segments} & 91.69\% \\
                $\hookrightarrow$ among "speech" segments & 99.50\% \\
                $\hookrightarrow$ among "music" segments & 85.45\% \\
                \vspace{-0.25cm} \\
                \underline{English segments} & 7.22\% \\
                $\hookrightarrow$ among "speech" segments & 0.30\% \\
                $\hookrightarrow$ among "music" segments & 12.75\% \\ 
                \bottomrule
            \end{tabular}
            \caption{Global statistics on the 100,000 hours audio dataset. $^*$Statistics obtained through heuristics and/or automatic tools.}
            \label{tab:global-dataset-stats}
        \end{table}

        In Table \ref{tab:global-dataset-stats} we show the global statistics of the 100,000 hours dataset we prepared.
        We obtain the speech and music segments counts with the following heuristics: segments are considered to have music if our music tool predicted less than 85\% of \textit{no-music}, and considered to be speech if more there is more than 20s according to InaSpeechSegmenter and less than 30\% of \textit{foreground-music}.

        The language is determined by Whisper. French segments represent 91.69\% of all segments according to Whisper. When filtering out segments not characterized as speech segments, French language represents 99.50\% of the dataset. On the other hand, speech segments with English only represent 0.30\% of them, while 12.75\% for music segments.

\begin{figure}[hbt!]
            \centering
            \includegraphics[width=\linewidth]{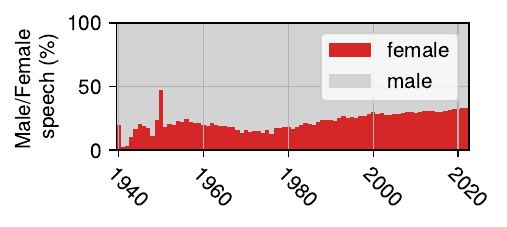}
            \vspace{-0.8cm}
            \caption{The per-gender speech ratio per year.
            }
            \label{fig:gender-speech-ratio-year}
        \end{figure}

        The gender balance is obtained with InaSpeechSegmenter, and is plotted per year in Figure \ref{fig:gender-speech-ratio-year}.
        The global speaking time for women in the whole 100,000h dataset is at 29.95\%.
    
\section{Self-Supervised Learning}
    \label{sec:ssl}
    We train audio SSL models following data2vec2 architecture presented by \citet{baevski2023efficient}. The architecture follows a teacher-student encoder setting, where the teacher corresponds to the exponentially moving average from the student weights. The student has to predict the masked audio sequence representation of the teacher.
    The encoder contains a CNN network, followed by 12 transformers layers. Our models contains 93.2M trainable parameters, similar to other speech models considered as \textit{base} models \citep{baevski2023efficient,lebenchmark2}.
    Models are trained for 100k steps with a cosine learning rate scheduler configured with a warmup of 8k steps to $7.5 \times 10^{-4}$. 
    The full training configuration and metrics are released along with the models.
    Each model is trained for around 70 GPU-hours on NVIDIA H100 80GB cards.

    Derived from the 100,000 hours training corpus presented in section \ref{sec:audio_datasets}, we define 6 pretraining datasets with controlled acoustic properties, in order to investigate the effects of data selection on downstream tasks.
    Each one of them is composed of 120,000 audio segments of 30s, for a total of 1,000h of speech.
    We train one model on each of these pretraining dataset. %
    The six models are:
    \begin{description}
    {\setlength{\itemsep}{0pt}
        \item[\base{}:] A model trained on a random sample of 1,000 hours.
        \item[\nomusic{}:] Trained on a subsample composed of segments detected as not containing music with our described heuristics.
        \item[\onlyspeech{}:] Trained on a subsample with only segments containing speech, as described by our heuristics.
        \item[\onlyfr{}:] Trained on a subsample on only French segments identified by Whisper.
        \item[\gender{}:] Trained on a subsample with a balanced proportion of male and female speech, as identified by InaSpeechSegmenter.
        With this subsample, we aim to see if these bias could be reduced through the pretraining dataset. 
        \item[\duplicates{}:] Trained on a subsample sourced from \textbf{\base{}}, where 1\% of the segments were duplicated 10 times. To have the same total duration as base, 10,800 other segments were randomly removed from this sample. The objective of this subsample is to create a comparable version to \texttt{\textbf{base}}, but without a deduplicating step, as presented in section \ref{sec:audio_dedup}.
    }
    \end{description}

\section{Evaluation on downstream tasks}
    
\begin{table*}[t]
    \centering

\begin{tabular}{l|c|ccccc}
\toprule
\textbf{Model} & \textbf{\textit{Global}} & \textbf{Antract} &        \textbf{QUAERO} &          \textbf{EPAC} &         \textbf{ESTER1} &        \textbf{REPERE} \\
\midrule
\textit{\small{}LB/wav2vec2-FR-1K-base} &           31.4 \textcolor{gray}{\small$\pm$0.1} &           28.4 \textcolor{gray}{\small$\pm$0.1} &           36.8 \textcolor{gray}{\small$\pm$0.5} &           29.8 \textcolor{gray}{\small$\pm$0.2} &           32.4 \textcolor{gray}{\small$\pm$0.2} &           34.0 \textcolor{gray}{\small$\pm$0.2} \\ \midrule
\texttt{base} & 15.3 \textcolor{gray}{\small$\pm$0.1} &           14.2 \textcolor{gray}{\small$\pm$0.1} &           18.5 \textcolor{gray}{\small$\pm$0.4} &           14.2 \textcolor{gray}{\small$\pm$0.2} &           15.8 \textcolor{gray}{\small$\pm$0.2} &           16.4 \textcolor{gray}{\small$\pm$0.2} \\
\texttt{no\_music}                    &  \textbf{14.4} \textcolor{gray}{\small$\pm$0.1} &           13.5 \textcolor{gray}{\small$\pm$0.1} &  \textbf{17.4} \textcolor{gray}{\small$\pm$0.4} &  \textbf{13.1} \textcolor{gray}{\small$\pm$0.2} &  \textbf{14.9} \textcolor{gray}{\small$\pm$0.2} &  \textbf{15.4} \textcolor{gray}{\small$\pm$0.2} \\
\texttt{only\_speech}                 &           \within{14.5} \textcolor{gray}{\small$\pm$0.1} &  \textbf{13.3} \textcolor{gray}{\small$\pm$0.1} &           18.0 \textcolor{gray}{\small$\pm$0.4} &           13.4 \textcolor{gray}{\small$\pm$0.2} &           \within{15.1} \textcolor{gray}{\small$\pm$0.2} &           \within{15.5} \textcolor{gray}{\small$\pm$0.2} \\
\texttt{only\_fr} &           15.0 \textcolor{gray}{\small$\pm$0.1} &           13.7 \textcolor{gray}{\small$\pm$0.1} &           18.2 \textcolor{gray}{\small$\pm$0.3} &           14.0 \textcolor{gray}{\small$\pm$0.2} &           15.5 \textcolor{gray}{\small$\pm$0.2} &           15.9 \textcolor{gray}{\small$\pm$0.2} \\
\texttt{gender} &           15.1 \textcolor{gray}{\small$\pm$0.1} &           14.0 \textcolor{gray}{\small$\pm$0.1} &           18.5 \textcolor{gray}{\small$\pm$0.4} &           13.9 \textcolor{gray}{\small$\pm$0.2} &           15.7 \textcolor{gray}{\small$\pm$0.2} &           16.1 \textcolor{gray}{\small$\pm$0.2} \\
\texttt{duplicates} &           15.6 \textcolor{gray}{\small$\pm$0.1} &           14.5 \textcolor{gray}{\small$\pm$0.1} &           18.9 \textcolor{gray}{\small$\pm$0.4} &           14.4 \textcolor{gray}{\small$\pm$0.2} &           16.2 \textcolor{gray}{\small$\pm$0.2} &           16.5 \textcolor{gray}{\small$\pm$0.2} \\
\bottomrule
\end{tabular}
    \caption{ASR results in WER ($\downarrow$) of the different models on the test sets of Antract, QUAERO, EPAC, ESTER1 and REPERE and globally across all corpora. Best results in \textbf{bold}, results within the confidence interval of the best model \within{underlined}.}
    \label{tab:asr-results}
\end{table*}

\begin{table}[htb!]
\centering
\begin{tabular}{lr|cc}
\toprule
\textbf{Dataset} &  & {\texttt{base}} & {\texttt{gender}} \\
\midrule
\multirow{3}{*}{\textbf{QUAERO}} & male &           \within{18.8} \textcolor{gray}{\small$\pm$0.4} &  \textbf{18.7} \textcolor{gray}{\small$\pm$0.4} \\
       & female &  \textbf{17.9} \textcolor{gray}{\small$\pm$0.6} &           \within{18.2} \textcolor{gray}{\small$\pm$0.6} \\ %
       & $\Delta_\text{rel}$ &                                           -4.9 &                                  \textbf{-2.7} \\ \midrule
\multirow{3}{*}{\textbf{EPAC}} & male &           14.0 \textcolor{gray}{\small$\pm$0.2} &  \textbf{13.7} \textcolor{gray}{\small$\pm$0.2} \\
       & female &           \within{15.8} \textcolor{gray}{\small$\pm$0.4} &  \textbf{15.2} \textcolor{gray}{\small$\pm$0.4} \\ %
       & $\Delta_\text{rel}$ &                                           12.1 &                                  \textbf{10.4} \\ \midrule
\multirow{3}{*}{\textbf{ESTER1}} & male &           \within{16.2} \textcolor{gray}{\small$\pm$0.2} &  \textbf{16.1} \textcolor{gray}{\small$\pm$0.2} \\
       & female &           \within{14.9} \textcolor{gray}{\small$\pm$0.3} &  \textbf{14.7} \textcolor{gray}{\small$\pm$0.3} \\ %
       & $\Delta_\text{rel}$ &                                  \textbf{-8.4} &                                           -9.1 \\
       \midrule
\multirow{3}{*}{\textbf{REPERE}} & male &           \within{16.4} \textcolor{gray}{\small$\pm$0.2} &  \textbf{16.1} \textcolor{gray}{\small$\pm$0.2} \\
       & female &           \within{16.9} \textcolor{gray}{\small$\pm$0.5} &  \textbf{16.5} \textcolor{gray}{\small$\pm$0.4} \\ %
       & $\Delta_\text{rel}$ &                                            3.0 &                                   \textbf{2.5} \\
\bottomrule
\end{tabular}
    \caption{Gendered ASR results in WER, and relative difference between male and female ($\Delta_\text{rel}$).}
    \label{tab:asr-gender-results}
\end{table}

    In this section we benchmark our audio encoders with multiple downstream tasks: automatic speech recognition, voice activity detection, music detection and speaker recognition. We also assess the ability of our models to recall their pretraining dataset with a membership inference attack. We compare our audio encoders with speech encoder baselines also trained on 1,000h of content.

    Unless otherwise stated, for all downstream tasks presented below, we train the downstream model and eventually finetune the audio encoder itself on either the official training subset of corpora used, or on our own split if it was not provided. We optimize hyperparameters on the development sets, and test on the remaining unseen data.
    We compute confidence intervals at 97.5\% using the bootstrap sampling strategy with $n=1000$.

    \subsection{Automatic Speech Recognition}
        \label{sec:asr}
        We benchmark the different audio encoder models on Automatic Speech Recognition (ASR), i.e. the task of transcribing speech.
        We feed the last transformer layer of the audio encoder into a linear projection layer. The model is trained with a Connectionist Temporal Classification (CTC) loss to predict character-level outputs.
        We first initialize this added layer by training it for 1k steps with the encoder freezed, and then we continue training with the rest of the model unfreezed for up to 30k steps.
        Models are evaluated with a greedy decoder and without a language model.

        ASR models are trained on the combination of Antract~\citelanguageresource{carrive_antract}, QUAERO~\citelanguageresource{boudahmane-etal-2011-advances}, EPAC~\citelanguageresource{esteve-etal-2010-epac}, ESTER1~\citelanguageresource{gravier-etal-2004-ester} and REPERE~\citelanguageresource{giraudel-etal-2012-repere} for a total of 258h, and evaluated on their test sets.
        We present in Table \ref{tab:asr-results} the results in Word Error Rate (WER) of the different audio encoders on this task.
        We compare our audio encoders with \textit{LB/wav2vec2-FR-1K-base} \citep{lebenchmark2}, also trained on 1,000 hours of French. Unlike our models, which are pretrained on spontaneous and diverse audio content, this model was pretrained on an audiobook dataset (MLS French \citealplanguageresource{Pratap2020MLSAL}).
        We notice our \base{} obtains much better results that this baseline model, with absolute an improvement of 16.1\% WER.
        Comparing our models trained on different subsamples, we can see that training without music (\nomusic{}) or with speech content only (\onlyspeech{}) improves the results compared to the standard \base{} setting.
        The model pretrained on data containing \duplicates{} seems to perform worse than it's \base{} counterpart. This confirms results previously observed by others \citep{lee-etal-2022-deduplicating}.

        In Table \ref{tab:asr-gender-results}, we present the per-gender WER on the datasets with speaker gender annotation: QUAERO, EPAC, ESTER1 and REPERE.
        Similar to \citet{zanonboito22_interspeech}, we compute the relative difference of WERs between male and female ($\Delta_\text{rel}$) as in Eq. \ref{eq:wer_male_female}.
        If the $\Delta_\text{rel}$ is greater than 0, the model is biased towards male as it performs better for male than for female, and better for female if the value is less than 0.
        Our gender analysis compares \base{} and \gender{} models as the two were trained with the same kind of content.
        During pretraining, 70\% of the speech for the \base{} model was from male, while it accounted to 54\% in the \gender{} model.

\begin{table*}[htb!]
\centering
    \resizebox{0.95\linewidth}{!}{
\begin{tabular}{l|cc|cccc}
\toprule
\multirow{3}{*}{\textbf{Model}} & \multicolumn{2}{c|}{\multirow{2}{*}{\textbf{\textit{Global}}}} & \textbf{generalist} & \textbf{generalist} & \textbf{music} & \textbf{news} \\
& & & \textbf{radio} & \textbf{tv} & \textbf{radio} & \textbf{tv} \\ \cmidrule(lr){2-3} \cmidrule(lr){4-4} \cmidrule(lr){5-5} \cmidrule(lr){6-6} \cmidrule(lr){7-7}
 & \textbf{Acc} & \textbf{F1} & \textbf{Acc} & \textbf{Acc} & \textbf{Acc} & \textbf{Acc} \\
\midrule
\textit{\small{}InaSpeechSegmenter}     &           93.0 \textcolor{gray}{\small$\pm$2.4} &           94.3 \textcolor{gray}{\small$\pm$2.0} &           95.4 \textcolor{gray}{\small$\pm$3.9} &           88.3 \textcolor{gray}{\small$\pm$4.1} &           \within{98.1} \textcolor{gray}{\small$\pm$3.1} &           94.9 \textcolor{gray}{\small$\pm$2.8} \\
\textit{\small{}pyannote}               &           88.7 \textcolor{gray}{\small$\pm$4.8} &           91.3 \textcolor{gray}{\small$\pm$4.2} &           \within{96.2} \textcolor{gray}{\small$\pm$3.0} &           89.6 \textcolor{gray}{\small$\pm$5.0} &          75.5 \textcolor{gray}{\small$\pm$16.8} &           96.1 \textcolor{gray}{\small$\pm$1.7} \\ \midrule
\textit{\small{}MFCC}                   &           89.9 \textcolor{gray}{\small$\pm$2.7} &           91.4 \textcolor{gray}{\small$\pm$2.9} &           91.3 \textcolor{gray}{\small$\pm$5.8} &           86.3 \textcolor{gray}{\small$\pm$4.6} &           92.2 \textcolor{gray}{\small$\pm$5.9} &           94.1 \textcolor{gray}{\small$\pm$2.9} \\
\textit{m-a-p/music2vec-v1}     &           \within{96.4} \textcolor{gray}{\small$\pm$1.1} &           \within{97.0} \textcolor{gray}{\small$\pm$1.0} &           \within{96.5} \textcolor{gray}{\small$\pm$2.7} &           \within{94.6} \textcolor{gray}{\small$\pm$2.1} &  \textbf{98.5} \textcolor{gray}{\small$\pm$1.7} &  \textbf{97.7} \textcolor{gray}{\small$\pm$1.2} \\
\textit{\small{}FB/data2vec-audio-base} &           95.2 \textcolor{gray}{\small$\pm$1.8} &           95.9 \textcolor{gray}{\small$\pm$1.7} &           \within{96.3} \textcolor{gray}{\small$\pm$3.3} &           93.2 \textcolor{gray}{\small$\pm$2.9} &           96.6 \textcolor{gray}{\small$\pm$4.0} &           \within{97.0} \textcolor{gray}{\small$\pm$2.6} \\
\midrule
\texttt{base}                   &  \textbf{96.8} \textcolor{gray}{\small$\pm$1.1} &  \textbf{97.3} \textcolor{gray}{\small$\pm$1.0} &  \textbf{97.6} \textcolor{gray}{\small$\pm$2.1} &  \textbf{95.6} \textcolor{gray}{\small$\pm$1.7} &           \within{97.8} \textcolor{gray}{\small$\pm$2.5} &  \textbf{97.7} \textcolor{gray}{\small$\pm$1.1} \\
\texttt{no\_music}               &  \within{96.0} \textcolor{gray}{\small$\pm$1.3} &           \within{96.6} \textcolor{gray}{\small$\pm$1.2} &           \within{96.6} \textcolor{gray}{\small$\pm$2.3} &           \within{94.4} \textcolor{gray}{\small$\pm$2.4} &           \within{97.6} \textcolor{gray}{\small$\pm$2.5} &           \within{97.1} \textcolor{gray}{\small$\pm$1.7} \\
\texttt{only\_speech}            &           \within{96.2} \textcolor{gray}{\small$\pm$1.2} &           \within{96.8} \textcolor{gray}{\small$\pm$1.2} &           \within{96.7} \textcolor{gray}{\small$\pm$2.3} &           \within{94.9} \textcolor{gray}{\small$\pm$2.0} &           \within{97.2} \textcolor{gray}{\small$\pm$3.2} &           \within{97.4} \textcolor{gray}{\small$\pm$1.1} \\
\texttt{only\_fr}                &           \within{96.3} \textcolor{gray}{\small$\pm$1.2} &           \within{96.9} \textcolor{gray}{\small$\pm$1.1} &           \within{96.5} \textcolor{gray}{\small$\pm$2.3} &           \within{95.1} \textcolor{gray}{\small$\pm$2.0} &           \within{97.6} \textcolor{gray}{\small$\pm$3.0} &           \within{97.4} \textcolor{gray}{\small$\pm$1.2} \\
\texttt{gender}                 &           \within{96.5} \textcolor{gray}{\small$\pm$1.3} &           \within{97.0} \textcolor{gray}{\small$\pm$1.2} &           \within{97.1} \textcolor{gray}{\small$\pm$2.1} &           \within{95.3} \textcolor{gray}{\small$\pm$1.9} &           \within{97.2} \textcolor{gray}{\small$\pm$3.5} &           \within{97.6} \textcolor{gray}{\small$\pm$1.1} \\
\texttt{duplicates}             &           \within{96.7} \textcolor{gray}{\small$\pm$1.0} &           \within{97.2} \textcolor{gray}{\small$\pm$1.0} &           \within{97.0} \textcolor{gray}{\small$\pm$2.3} &           \within{95.4} \textcolor{gray}{\small$\pm$2.0} &  \textbf{98.5} \textcolor{gray}{\small$\pm$1.6} &           \within{97.4} \textcolor{gray}{\small$\pm$1.3} \\
\bottomrule
\end{tabular}
}
    \caption{Accuracy, and F1-Score of the different embedding representations models for the VAD task on the test set of InaGVAD (both globally and per channel category). Best results in \textbf{bold}, results within the confidence interval of the best model \within{underlined}.}    
    \label{tab:vad}
\end{table*}

        \begin{align}
            \label{eq:wer_male_female}
            \Delta_\text{rel} = 100 \times \frac{\text{WER}_\text{F} - \text{WER}_\text{M}}{0.5 \times (\text{WER}_\text{F} + \text{WER}_\text{M})}
        \end{align}

        Globally, we can see EPAC women transcriptions seem more difficult than men's. On the other hand, we notice QUAERO and ESTER seems easier for female speech.
        This result could be explained by biased speaker roles inside broadcast content corpora \citep{addadecker05_interspeech,genderasr}.
        Regarding the model gender bias, measured with the relative difference of WERs ($\Delta_\text{rel}$), we notice that \gender{} obtains less bias than \base{} for all datasets except ESTER1, which indicates training on balanced gender corpora could be beneficial. However, it should be noted that the overall differences between \base{} and \gender{} results are not statically different.
    
    \subsection{Voice Activity Detection}
        \label{sec:vad}
        We evaluate the impact of the pretraining set on a simple Voice Activity Detection downstream task. VAD aims to identify and segment portions of an input audio recording that contain human speech, separating them from silence, breathing, background noise or music. For our experiments, it is treated as a binary frame classification problem with two mutually exclusive classes (speech and non speech). 

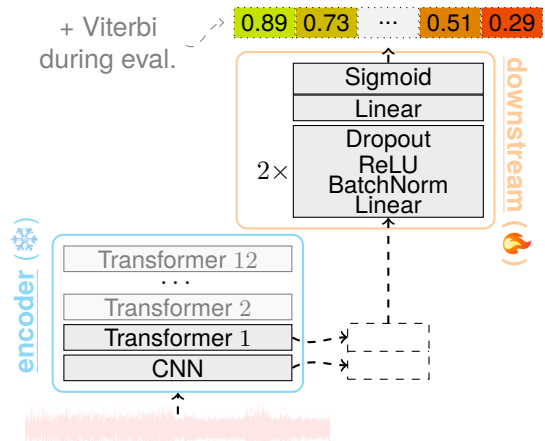
\begin{figure}[htb!]
    \centering
\begin{tikzpicture}[font=\sffamily, every node/.style={align=center}]

\node[] at (0, 0) (sound1) {\includegraphics[width=4cm, trim={10cm 0 6cm 0},clip]{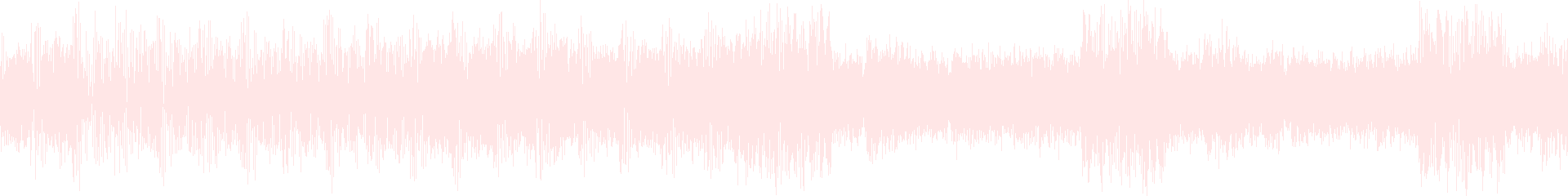}};

\node[transformerLayer, above=0.15cm of sound1] (cnn) {CNN};
\node[transformerLayer, above=0.05cm of cnn] (t1) {Transformer $1$};
\node[transformerLayerIgnored, above=0.05cm of t1] (t2) {Transformer $2$};
\node[draw=none, above=0.0cm of t2] (dots) {$\dots$};
\node[transformerLayerIgnored, above=0.0cm of dots] (t12) {Transformer $12$};
\node[draw, rounded corners, thick, fit=(cnn)(t1)(t2)(t12), inner sep=4pt, color=cyan!40, label={[rotate=90, yshift=1em, xshift=4em, color=cyan!40]left:\textbf{\underline{encoder}} (\emoji{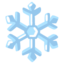})}] (encoder) {};

\node[draw, fill=gray!15, minimum width=2.5cm, minimum height=0.2cm, above right=0.3cm of encoder.north, xshift=1.3cm,execute at begin node=\setlength{\baselineskip}{2pt}, inner sep=1.5pt] (l1) {\small{}Dropout\\\small{}ReLU\\\small{}BatchNorm\\\small{}Linear};
\node[left=0.1cm of l1, xshift=0.2cm] (times) {$2\times$};
\node[draw, fill=gray!15, minimum width=2.5cm, minimum height=0.15cm, above=0.05cm of l1, inner sep=1.5pt] (l10) {\small{}Linear};
\node[draw, fill=gray!15, minimum width=2.5cm, minimum height=0.15cm, above=0cm of l10, inner sep=1.5pt] (l11) {\small{}Sigmoid};

  \matrix [above=0.2cm of l11,] (output2)
  {
    \node[draw, dotted, fill=lime!89!red, minimum width=0.8cm, minimum height=0.40cm, inner sep=2pt] {\footnotesize{}0.89}; &
    \node[draw, dotted, fill=lime!73!red, minimum width=0.8cm, minimum height=0.40cm, inner sep=2pt] {\footnotesize0.73}; &
    \node[draw, dotted, fill=gray!10,     minimum width=0.8cm, minimum height=0.40cm, inner sep=2pt] {\footnotesize...}; &
    \node[draw, dotted, fill=lime!51!red, minimum width=0.8cm, minimum height=0.40cm, inner sep=2pt] {\footnotesize0.51}; &
    \node[draw, dotted, fill=lime!29!red, minimum width=0.8cm, minimum height=0.40cm, inner sep=2pt] {\footnotesize0.29}; \\
  };

\node[left=0.5cm of output2, text=gray, yshift=-0.3cm] (viterbi) {+ Viterbi\\ during eval.};
\draw (viterbi.east) edge[out=-30,in=180,->,draw=gray,dashed] (output2.west);

\node[minimum width=1cm, minimum height=0.3cm, below=1.4cm of l1, outer sep=1pt, inner sep=0pt] (hidden1) {};
\node[minimum width=1cm, minimum height=0.3cm, below=0cm of hidden1.south, anchor=north, outer sep=1pt, inner sep=0pt] (hidden2) {};
\node[fit=(hidden1)(hidden2), draw, dashed, inner sep=0pt, outer sep=0pt] (hidden) {};
\draw[dashed, -] (hidden1.south west) -- (hidden1.south east);

\node[draw, rounded corners, thick, fit=(l1)(times)(l10)(l11), inner sep=4pt, color=orange!40, label={[rotate=-90, yshift=8pt, xshift=-3.5em, color=orange!40]right:\textbf{\underline{downstream} (\emoji{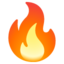})}}] (dowstream) {};

\draw[->, thick, dashed] (sound1.center) ++(0,0.2) -- (encoder.south);

\draw[->, thick, dashed] (l11.north) -- (output2.south);

\draw (cnn.east) edge[out=30,in=180,->,dashed, thick] (hidden2.west);
\draw (t1.east) edge[out=-30,in=180,->,dashed, thick] (hidden1.west);

\draw[->, thick, dashed] (hidden.north) -- (hidden.north |- l1.south) -- (l1.south);

\end{tikzpicture}
    \vspace{-0.25cm}
\caption{Overview of architecture used of the Voice Activity Detection and Music Detection models.}
    \label{fig:vad-music-detection-model}
\end{figure}

        Our downstream models uses the concatenation of the representations extracted from the CNN and the first transformer block of the SSL model.
        Based on empirical results, using only these two layers provide a good tradeoff between accuracy and inference speed.
        The concatenated hidden representations are frozen and then fed to a simple MLP classifier, described in Figure \ref{fig:vad-music-detection-model}.
        We predict classes on frames sampled at 50Hz, aligned with the hidden representations frequency, and use the Viterbi algorithm for smoothing transition probabilities of 1\% for \textit{speech} to \textit{non speech} and inversely.

        Models are trained on the \textit{dev}\footnote{We sample it into training (80\%) and validation sets.} subset of InaGVAD \citelanguageresource{doukhan-etal-2024-inagvad}, a corpus designed to evaluate VAD systems on French audiovisual broadcast content.

        The global and per-category results on InaGVAD are presented in Table \ref{tab:vad}.
        We compare our downstream models with previous state-of-the-art VAD baselines: InaSpeechSegmenter \citep{doukhan_inaspeechsegmenter} and pyannote.audio \citep{bredin21_interspeech}.
        While earlier systems struggled with generalist TV and music radio content, our trained models show a clear improvement and strong performances across all categories. Previous systems were typically trained on separate speech and music audio files,  whereas incorporating a more challenging broadcast type during training proved beneficial for our models. Overall, the \base{} model performs best, although the music2vec encoder obtains good results as well. %

    \subsection{Music Detection}
        \label{sec:music}

\begin{table*}[htbp!]
    \centering
\begin{tabular}{l|ccc|ccc}
\toprule
\multirow{3}{*}{\textbf{Model}} & \multicolumn{3}{c|}{\textbf{\textit{Global}}} &      \textbf{Mirex2015} & \textbf{OpenBMAT} & \textbf{Seyerlehner} \\ \cmidrule(lr){2-4} \cmidrule(lr){5-5} \cmidrule(lr){6-6} \cmidrule(lr){7-7}
{} & \textbf{F1} & \textbf{P} & \textbf{R} & \textbf{F1} & \textbf{F1} & \textbf{F1} \\
&  \textcolor{gray}{\footnotesize$\pm$0.1} &  \textcolor{gray}{\footnotesize$\pm$0.1} &  \textcolor{gray}{\footnotesize$\pm$0.1} &  \textcolor{gray}{\footnotesize$\pm$0.1} &  \textcolor{gray}{\footnotesize$\pm$0.1} &  \textcolor{gray}{\footnotesize$\pm$0.2} \\
\midrule
\textit{\small{}MFCC}                          &                             82.8 &                             76.3 &                             90.5 &                             95.4 &                             79.8 &                             83.4 \\
\textit{\small{}m-a-p/music2vec-v1}         &                    \textbf{91.2} &                    \textbf{90.6} &                    \textbf{91.8} &                             96.7 &                    \textbf{89.7} &                    \textbf{92.0} \\
\textit{\small{}FB/data2vec-audio-base}  &                             87.5 &                             84.6 &                             90.6 &                             97.0 &                             85.4 &                             87.7 \\
\midrule
\texttt{base}                          &                             89.4 &                             89.4 &                             89.3 &                    \textbf{97.3} &                             87.1 &                             91.0 \\
\texttt{no\_music}                     &                             87.1 &                             86.5 &                             87.7 &                             96.9 &                             84.5 &                             88.4 \\
\texttt{only\_speech}                  &                             87.5 &                             87.2 &                             87.9 &                             97.0 &                             85.0 &                             88.9 \\
\texttt{only\_fr}                      &                             87.7 &                             84.2 &                             91.6 &                             \within{97.2} &                             85.1 &                             89.8 \\
\texttt{gender}              &                             88.3 &                             86.2 &                             90.5 &                             \textbf{97.3} &                             85.6 &                             90.7 \\
\texttt{duplicates}           &                             88.9 &                             88.4 &                             89.4 &                             \within{97.2} &                             86.5 &                             90.8 \\
\bottomrule
\end{tabular}
    \caption{Frame-level F1-Score (F1), Precision (P) and Recall (R) for the music detection task
    globally and individually on each dataset.
    The confidence interval is the maximum of all models for a given column. %
    }
    \label{tab:music-detection-global}
\end{table*}

        Next, we evaluate our models on a music segmentation task. Given an input audio file, the goal of this task is to determine timestamps of segments containing music.
        As with VAD, we configure our downstream model to use the concatenation CNN representations as well as hidden representations of the 1$^\text{st}$ transformer block. This concatenation is fed into a MLP classifier, similar to the one used for VAD, described in Figure \ref{fig:vad-music-detection-model}.
        We predict a binary class (music/no-music) at a 50Hz frequency.
        We use the Viterbi algorithm with transition probabilities of $5\%$ for \textit{music} to \textit{no-music} and vice-versa.
        To manage the memory consumption during training and evaluation, we slice audio files into chunks of 30s, with no overlap.

        Our downstream models are trained on OpenBMAT \citelanguageresource{MelndezCataln2019}, the Music/Speech detection dataset of the Mirex 2015 competition \citelanguageresource{mirex2015}, and the music detection corpus presented by \citetlanguageresource{seyerlehner2007automatic}, denoted as Seyerlehner.

        We present in Table \ref{tab:music-detection-global} our results. We can see that the music2vec-v1 pretrained model, presented by \citet{li2022mapmusic2vecsimpleeffectivebaseline} achives the best overall results. This model has been pretrained on 1,000 hours of music audio content, and is therefore only suitable for MIR tasks.
        Next we observe that our models obtain the best results when pretrained on content which includes music and non-speech: \base{}, \gender{} and \duplicates{}. The model using \nomusic{}, pretrained on content without music has a F1-Score 2.2\% lower than it's \base{} counterpart.
        
\begin{table*}[htb!]
	\centering
    \resizebox{0.95\linewidth}{!}{
	\begin{tabular}{l|ccc|ccc|ccc}
		\toprule
		\multirow{2}{*}{\textbf{Model}} & \multicolumn{3}{c|}{\textbf{VoxCeleb1-O}} & \multicolumn{3}{c|}{\textbf{VoxCeleb1-E}} & \multicolumn{3}{c}{\textbf{VoxCeleb1-H}}\\
		\cmidrule(lr){2-4} \cmidrule(lr){5-7} \cmidrule(lr){8-10}  
		& \textbf{EER} & \textbf{minDCF1} & \textbf{minDCF5} & \textbf{EER} & \textbf{minDCF1} & \textbf{minDCF5} & \textbf{EER} & \textbf{minDCF1} & \textbf{minDCF5} \\
	   \midrule
       \textit{\small{}MS/WavLM-base} & \textbf{6.32}  & \textbf{0.540} & \textbf{0.397} & \textbf{6.98} & \textbf{0.575} & \textbf{0.400} & \textbf{11.45} & \textbf{0.707} & \textbf{0.553} \\ \midrule
 		\texttt{base}                &  8.20  &  0.591 & 0.450 &  8.02 & 0.615 & 0.455 &  11.97 & 0.722 & 0.572  \\ 
		\texttt{only\_speech}  & 8.00 & 0.634 & 0.454 &   7.84 & 0.612 & 0.440 & {11.50} & {0.710} & {0.561} \\  
		\texttt{gender}                & {7.35}      &  {0.576}    & {0.409}   & {7.83}   & {0.610}  & {0.439} & 11.64 & 0.712  & 0.567 \\
		\bottomrule
	\end{tabular}}
	\caption{EER(\%) and minDCF of pretrained SSL models on the VoxCeleb Speaker Recognition benchmark.}    
	\label{tab:sr}
\end{table*}

    \subsection{Speaker Recognition}
        \label{sec:sr}
        We evaluate the representations extracted by our models (\base{}, \onlyspeech{}, and \gender{}) and \textit{\small{}MS/WavLM-base} \citep{sanyuan22_stsp} using the VoxCeleb speaker recognition benchmark \citep{voxceleb1,voxceleb2}.
        To aggregate the representation layers, we use the attention-based MHFA backend described in \citet{peng23_slt}. The 12 transformer layers and the CNN layer are combined with the attention-based model to produce 256-dimensional embedding. The cosine distance is used to compute the test score between the test and the target embeddings. Only the backend model parameters were finetuned during 50 epochs using a subset of 1,000 speakers derived from the VoxCeleb2 development set with AAM-softmax, a margin of 0.2, and a scaling of 30.
        Both Equal Error Rate (EER) and the minimum Detection Cost Function (minDCF) are used to evaluate the
        speaker verification systems. The target probability $P_{tar}$ is set to $0.01$ or $0.05$, for DCF1 and DCF5, respectively. $C_{fa}$ and $C_{fr}$ have an equal weight of $1.0$.

        As shown in Table \ref{tab:sr}, the performances of all evaluated models are consistent. The \gender{} model achieves the best overall scores.
        This suggests that speaker identity information can be inferred from the hidden layer representations of our SSL model.
        WavLM achieves slightly better performance, likely due to its pretraining on English speech, which matches the language used in the VoxCeleb dataset. However, these scores remain below those usually reported by finetuned SSL models, indicating that finetuning SSL models is necessary to achieve higher performance, as finetuned models adapt their representations more effectively to the target application. During training, analysis of the learned weights of the backend model reveals higher weights for the first layers, a pattern that aligns with previous work \citep{sanyuan22_stsp}.
        This indicates that speaker identity features are represented in the first layers.
        
    \subsection{Membership Inference Attack}
        \label{sec:mia}
        We evaluate the ability of the audio encoder to remember its pretraining dataset by performing a Membership Inference Attack (MIA).
        Our intuition behind this attack is to assess whether the encoder is able to memorize some data examples seen during the pretraining. While a model remembering it pretraining dataset would not necessarily enable its extraction or generation, a model not even able to remember its pretraining data would lower this risk. 
        
        First, we construct a 22-hour long training dataset, composed of 1320 segments seen once during the pretraining of \base{} and \duplicates{}, and 1320 segments never seen in any of them.
        
\begin{figure}
    \centering
\begin{tikzpicture}[font=\sffamily, every node/.style={align=center}]

\node[] at (0.5, 0.2) (sound2) {\includegraphics[width=4cm, trim={10cm 0 6cm 0},clip]{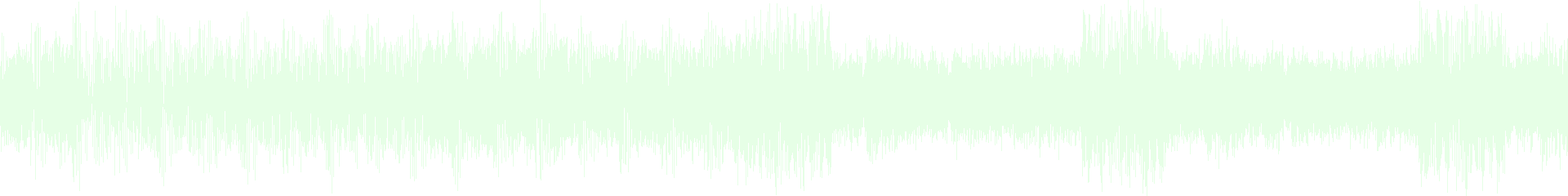}};
\node[] at (0, 0) (sound1) {\includegraphics[width=4cm, trim={10cm 0 6cm 0},clip]{figures/bb36a818-51b0-47f4-a8fb-54cfaf9976dd.png}};

\node[transformerLayer, above=0.25cm of sound1] (cnn) {CNN};
\node[transformerLayer, above=0.05cm of cnn] (t1) {Transformer $1$};
\node[transformerLayer, above=0.05cm of t1] (t2) {Transformer $2$};
\node[draw=none, above=0.0cm of t2] (dots) {$\dots$};
\node[transformerLayer, above=0.0cm of dots] (t12) {Transformer $12$};
\node[draw, rounded corners, thick, fit=(cnn)(t1)(t2)(t12), inner sep=4pt, color=cyan!40, label={[rotate=90, yshift=1em, xshift=4em, color=cyan!40]left:\textbf{\underline{encoder}} (\emoji{snowflake})}] (encoder) {};

\node[draw, fill=gray!15, minimum width=2.5cm, minimum height=0.2cm, above right=1.0cm of encoder.north, xshift=0.7cm, execute at begin node=\setlength{\baselineskip}{1.5pt}, inner sep=1.5pt] (l1) {\small{}Dropout\\\small{}ReLU\\\small{}BatchNorm\\\small{}Linear};
\node[left=0.1cm of l1, xshift=0.2cm] (times) {$2\times$};
\node[draw, fill=gray!15, minimum width=2.5cm, minimum height=0.15cm, above=0.05cm of l1, inner sep=1.5pt] (l10) {\small{}Linear};
\node[draw, fill=gray!15, minimum width=2.5cm, minimum height=0.15cm, above=0cm of l10, inner sep=1.5pt] (l11) {\small{}Sigmoid};

\node[draw, minimum height=0.2cm, above=0.50cm of l11, xshift=0.45cm, fill=green!10, rounded corners, inner sep=2pt] (output1) {\footnotesize{}0.21};
\draw[->, thick, dashed, color=green!50!black] (l11.north -| output1.south) -- (output1.south);
\node[draw, minimum height=0.2cm, above=0.25cm of l11, fill=red!10, rounded corners, inner sep=2pt] (output2) {\footnotesize{}0.89};
\node[circle, draw, minimum size=0.35cm, inner sep=0pt, below=0.15cm of l1, thick] (circle) {};

\draw[line width=0.8pt] (circle.north west) -- (circle.south east);
\draw[line width=0.8pt] (circle.south west) -- (circle.north east);

\node[draw, rounded corners, thick, fit=(l1)(times)(l10)(l11)(circle), inner sep=4pt, color=orange!40, label={[rotate=90, yshift=1em, xshift=5em, color=orange!40]left:\textbf{\underline{downstream} (\emoji{fire})}}] (dowstream) {};

\draw[->, thick, dashed, color=red!50!black] (sound1.center) ++(0,0.2) -- (encoder.south);
\draw[->, thick, dashed, color=green!50!black] (sound2.center) ++(0,0.2) -- (sound2.center |- encoder.south);

\draw[->, thick, dashed, color=red!50!black] (l11.north) -- (output2.south);

\draw[->, thick, dashed] (cnn.east) -- (cnn.east -| circle.south) -- (circle.south);
\draw[-, thick, dashed] (t1.east) -- (t1.east -| circle.south);
\draw[-, thick, dashed] (t2.east) -- (t2.east -| circle.south);
\draw[-, thick, dashed] (t12.east) -- (t12.east -| circle.south);

\draw[->, thick, dashed] (circle.north) -- (circle.north |- l1.south) -- (l1.south);

\end{tikzpicture}
    \vspace{-0.25cm}
\caption{Overview of the proposed Membership Inference Attack (MIA) downstream model.}
    \label{fig:mia-model}
\end{figure}

        The MIA downstream model is described in Figure \ref{fig:mia-model}. It is composed of a weighted sum of hidden states (CNN+transformers layers), followed by a MLP. The weights of the weighted sum are learned along with the linear layers. We freeze the audio encoder during the training of the downstream model. %

        During training, we notice the downstream MIA model struggles to converge on the development set, exhibiting the difficulty of the task.
        
        Next, for evaluating the MIA, we construct three different test sets of 10 hours (1,200 segments each): \textit{unseen} where neither of the model has seen the examples during their pretraining ; \textit{once} where both models saw the examples only once in the pretraining dataset ; \textit{duplicated} where the \duplicates{} model saw the examples 10 times in its pretraining dataset.
        Neither of the three sets overlaps with each other nor the 22h downstream training set.

\begin{figure}
            \centering
            \includegraphics[width=0.75\linewidth]{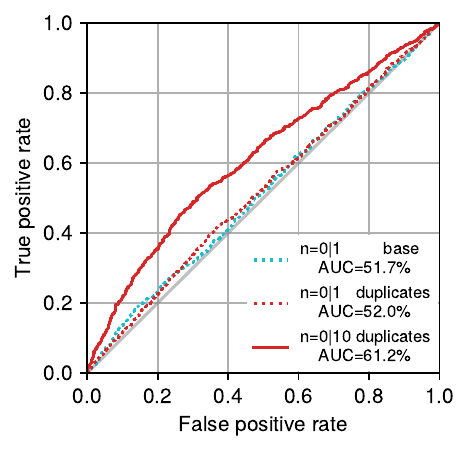}
            \vspace{-0.5cm}
            \caption[Caption for LOF]{ROC curves for the membership inference attack.}
            \label{fig:roc-membership}
        \end{figure}

        We present in Figure \ref{fig:roc-membership} the Receiver Operating Characteristic (ROC) curves\footnote{For the sake of clarity, the ROC curve for the \base{} model on $n=0|10$ is not shown. It has an AUC of 50.5\%.} for the \base{} and \duplicates{} models on \textit{unseen} vs \textit{once} ($n=0|1$) and \textit{unseen} vs \textit{duplicated} ($n=0|10$).
        
        We can see that when testing on examples seen multiple times during pretraining ($n=0|10$ for \duplicates{} model), the attack is able to succeed better than by chance, with an Area Under the Curve (AUC) of 61.2\%.
        The membership inference attack on \base{} fail, with an AUC of 51.7\% suggesting the model was not able to remember its pretraining dataset.
        We notice that the \duplicates{} model was not able to remember examples seen once ($n=0|1$) with an AUC of 52.0\%.
        Overall, this suggests that pretraining with duplicates elements do not make unduplicated segments more recoverable, contrary to duplicated elements, which become recoverable with an AUC of 61.2\%.

\section{Conclusion}
    In this paper, we construct a 100,000 hours pretraining corpus of audiovisual TV and Radio content. We build 6 pretrained audio SSL models that we benchmark on various downstream evaluations.

    Our observations shows that for speech recognition, pretraining on content without music improves the results compared to more diverse content. Gender-wise, pretraining on as much male than female speech seems to reduce the gender bias between male and female WER.
    On voice activity detection, pretraining on diverse audiovisual data obtains the best results.
    Regarding music detection, we observe that pretraining purely on music achieves the best results, although training on various audiovisual content obtains close results.
    For speaker recognition, our models achieved performance comparable to another speech SSL encoder, with the Gender-wise model achieving the best results.
    Overall, across all tasks, we can say that the results between downstream models are close enough to the specialized models that we can consider pretraining a general purpose audio model on the whole 100,000 hours corpus, suitable for both music and speech tasks. This experiment is left for future works.
    Lastly, the membership attack highlighted the importance of data deduplication, with a great reduction of training data memorization.

\section{Acknowledgments}
The authors would like to thank Jean-Hugues Chenot, Nicolas Hervé and Sandrine Depoix for their help during the construction of the INA dataset and with the audio deduplication tool.
They also thank Aude Formagne regarding the legal challenges of publishing the pretrained models. %

This research has been funded by the French National Research Agency (ANR), project ``Pantagruel'' (ANR-23-IAS1-0001).
This work was performed using HPC resources from GENCI at IDRIS and CINES under the allocations 2022-A0131013801, 2023-A0151013801, 2024-A0171013801, 2024-A0161015074, and 2025-A0191013801 on the Jean Zay and Adastra supercomputers.

\section{Bibliographical References}\label{sec:reference}

\bibliographystyle{lrec2026-natbib}
\bibliography{lrec2026-example}

\section{Language Resource References}
\label{lr:ref}
\bibliographystylelanguageresource{lrec2026-natbib}
\bibliographylanguageresource{languageresource}

\end{document}